\documentclass[12pt]{mn2e}
\usepackage{graphicx}
\usepackage{epsf}
\usepackage{lscape}
\usepackage{longtable}
\usepackage{rotating}
\usepackage{color}
\usepackage{xspace}

\newcommand{\aap}{A\&A}

\newcommand{\apj}{ApJ}

\newcommand{\mnras}{MNRAS}

\begin{document}
\topmargin -0.5in 

\title[The intrinsic UV properties of star forming galaxies]{Predictions for the intrinsic
UV continuum properties of star forming galaxies and the implications for inferring dust extinction}

\author[Stephen M. Wilkins, et al.\ ]  
{
Stephen M. Wilkins$^{1}$\thanks{E-mail: stephen.wilkins@physics.ox.ac.uk}, Violeta Gonzalez-Perez$^{2}$, Cedric G. Lacey$^{2}$, Carlton M. Baugh$^{2}$ \\
$^1$\,University of Oxford, Department of Physics, Denys Wilkinson Building, Keble Road, OX1 3RH, U.K. \\
$^{2}$\,Institute for Computational Cosmology, Department of Physics, University of Durham, South Road, Durham, DH1 3LE, U.K.\\
}
\maketitle 

\begin{abstract}
The observed ultraviolet continuum (UVC) slope is potentially a powerful diagnostic of dust obscuration in star forming galaxies. However, the intrinsic slope is also sensitive to the form of the stellar initial mass function (IMF) and to the recent star formation and metal enrichment histories of a galaxy. Using the {\sc galform} semi-analytical model of galaxy formation, we investigate the intrinsic distribution of UVC slopes. For star-forming galaxies, we find that the intrinsic distribution of UVC slopes at $z=0$, parameterised by the power law index $\beta$, has a standard deviation of $\sigma_{\beta}\simeq 0.30$. This suggests an uncertainty on the inferred UV attenuation of $A_{\rm fuv}\simeq 0.7$ (assuming a Calzetti attenuation curve) for an individual object, even with perfect photometry. Furthermore, we find that the intrinsic UVC slope correlates with star formation rate, intrinsic UV luminosity, stellar mass and redshift. These correlations have implications for the interpretation of trends in the observed UVC slope with these quantities irrespective of the sample size or quality of the photometry. Our results suggest that in some cases the attenuation by dust has been incorrectly estimated.  
\end{abstract}


\section{Introduction}

The ultraviolet continuum (UVC) slope has been proposed as a robust diagnostic of dust attenuation in star forming galaxies (e.g. Meurer et al. 1999). The usefulness of the UVC slope is most apparent at high-redshift where alternative diagnostics of dust attenuation such as the far-IR are generally inaccessible and galaxies are selected on the basis of their rest-frame UV emission (e.g. Wilkins et al. 2010; Wilkins et al. 2011a). This has prompted several studies which carefully study the UVC slopes of high-redshift galaxies in an attempt to constrain the degree of dust obscuration in UV selected samples (e.g. Bouwens et al. 2009, Dunlop et al. 2011, Wilkins et al. 2011b, Bouwens et al. 2011, Finkelstein et al. 2012). 

In all of these studies a simple linear relation is assumed between the observed UVC slope and the amount of dust attenuation. This requires the assumption of a unique {\em intrinsic} UVC slope. However, the intrinsic UVC slope is also affected by a number of other properties besides dust, including the recent star formation and metal enrichment histories, and the form of the stellar initial mass function (IMF). The sensitivity of the intrinsic slope to these properties potentially limits the usefulness of the observed UVC slope as an accurate diagnostic of dust attenuation. In this work we employ the {\sc galform} semi-analytical model of galaxy formation (first developed by Cole et al. 2000) to produce realistic star formation and metal enrichment histories with which to determine both the intrinsic distribution of UVC slopes $\beta_{i}$ and to investigate how this distribution varies with stellar mass, star formation rate, UV luminosity and redshift. Lacey et al. (2011) have shown this model can reproduce the observed UV luminosity function over a wide range of redshifts (see also Gonzalez-Perez et al. 2012, {\em in prep}).

This paper is organised as follows. In \S\ref{sec:i.galform} and \S\ref{sec:i.filters} we introduce the {\sc galform} semi-analytical galaxy formation model and filter convention respectively. In Section \ref{sec:pp} we discuss the principal physical properties which affect the observed UVC and investigate, using the {\sc pegase.2} stellar population synthesis (SPS) model (Fioc \& Rocca-Volmerange 1997, Fioc \& Rocca-Volmerange 1999), how the UVC slope is affected by dust and changes in the recent star formation and metal enrichment histories, and by the choice of IMF. In Section \ref{sec:pp} we also investigate the distribution of UVC colours/slopes predicted by the {\sc Galform} model as a result of the predicted variety in the star formation history (SFH) alone (\S\ref{sec:pp.sfh}) and in the metal enrichment history (\S\ref{sec:pp.Z}). In Section \ref{sec:a} we investigate how the distribution of intrinsic UVC slopes is affected by the star formation rate (\S\ref{sec:a.sfr}), intrinsic UV luminosity (\S\ref{sec:a.L_fuv}), stellar mass (\S\ref{sec:a.smass}) and redshift (\S\ref{sec:a.z}). In Section \ref{sec:c} we present our conclusions.

\section{Modelling Approach}

In this section we give an overview of the {\sc galform} galaxy formation model (\S\ref{sec:i.galform}) and set out how the UV continuum slope is measured (\S\ref{sec:i.filters}).

\subsection{The Galaxy Formation Model}\label{sec:i.galform}

We predict the intrinsic UV-slope of galaxies in a $\Lambda$CDM universe using the {\sc galform} semi-analytical galaxy formation model developed by Cole et al. (2000). Semi-analytical models use physically motivated equations to follow the fate of baryons in a universe in which structure grows hierarchically through gravitational instability (see Baugh 2006 for an overview of hierarchical galaxy formation models). 

{\sc galform} follows the main processes which shape the formation and evolution of galaxies. These include: (i) the collapse and merging of dark matter haloes; (ii) the shock-heating and radiative cooling of gas inside dark matter haloes, leading to the formation of galaxy discs; (iii) quiescent star formation in galaxy discs; (iv) feedback from supernovae, from active galactic nuclei (AGN) and from photoionization of the intergalactic medium (IGM); (v) chemical enrichment of stars and gas; (vi) galaxy mergers driven by dynamical friction within common dark matter haloes, leading to the formation of stellar spheroids, which also may trigger bursts of star formation. The end product of the calculation is a prediction for the number and properties of galaxies which reside within dark matter haloes of different masses. The outputs include the star formation and metal enrichment histories for each galaxy, including the contribution of merger driven starbursts. The star formation histories predicted by the model are in general more complex than the exponentially decaying models typically assumed to interpret observations (see Baugh 2006 for examples).

In this paper we focus our attention on the Baugh et al. 2005 (hereafter B05) model. Some of the key features of this model are (i) a time-scale for quiescent star formation that is assumed to vary as a power of the disc circular velocity (see Lagos et al. 2011 for a study of different star formation laws in quiescent galaxies), (ii) bursts of star formation are triggered only by galaxy mergers, (iii) the {\em default} implementation of this model adopts a Kennicutt (1983) IMF ($\xi={\rm d}N/{\rm d}m\propto m^{-2.5}$ for $m>1\,{\rm M_{\odot}}$ and $\xi\propto m^{-1.4}$ for $m<1\,{\rm M_{\odot}}$, c.f. Salpeter 1955: $\xi\propto m^{-2.35}$) in quiescent star formation in galactic discs, while in starbursts a top-heavy IMF ($\xi\propto m^{-1}$) is assumed, (iv) the inclusion of supernova feedback with superwinds (see Benson et al. 2003 for a discussion of the effect that feedback has on the luminosity function of galaxies), and (v) the reionisation of the intergalactic medium is approximated by a simple model in which gas cooling is completely suppressed in haloes with circular velocities less than $30\,{\rm km s^{-1}}$ at redshifts $z<10$ (Lacey et al. 2011). The parameters of this model were fixed with reference to a subset of the available observations of galaxies, mostly at low redshift. The B05 model uses, by default, the simple stellar population (SSP) spectral energy distributions (SEDs) generated by Bressan, Granato, \& Silva (1998), using the Padova 1994 stellar evolution tracks and the model stellar atmospheres from Kurucz (1993)\footnote{These are the same isochorones and stellar atmospheres used by Bruzual \& Charlot (2003).}. The B05 model uses the canonical ($\Lambda$CDM) parameters: matter density, $\Omega_{0}=0.3$, cosmological constant, $\Omega_{\Lambda} = 0.7$, baryon density, $\Omega_{b}=0.04$, a normalisation of density fluctuations given by $\sigma_{8}=0.93$ and a Hubble constant today $H_0=70$ km$\,{\rm s}^{-1}$Mpc$^{-1}$. The B05 model employs merger trees generated using the Monte Carlo algorithm introduced by Parkinson et al. (2008). The B05 model also includes a dust model in which stars and dust are mixed together. The dust extinction in {\sc galform} and the impact on the rest-frame UV is discussed in Lacey et al. (2011) and Gonzalez-Perez et al. (2012, {\em in prep}). In the calculations presented here we study the intrinsic UVC predicted by {\sc galform}, and so omit dust extinction in the model predictions.

In addition to reproducing local galaxy data, the B05 model matches the number and redshift distribution of galaxies detected by their emission at sub-millimetre wavelengths, the luminosity function of Lyman break galaxies and the abundance and clustering of Lyman-alpha emitters (Orsi et al. 2008). No parameters have been tuned for the study presented here. We refer the reader to B05 and Lacey et al. (2008, 2011) for a full description of this model. As a method of exploring differences in the distribution of UVC slopes arising from the initial mass function, star formation history and metal enrichment history we consider 3 seperate implementations of the B05 model:
\begin{itemize}
\item {\em default} - the default implementation of the model as described by B05.
\item {\em single IMF} - same as the default implementation except the top-heavy IMF in starbursts is replaced by the Kennicutt IMF, so that this IMF is used in all modes of star formation.
\item {\em single IMF + single metallicity} - same as the default implementation except the top-heavy IMF in starbursts is replaced by the Kennicutt IMF and all stars are assumed to have the same metallicity ($Z=Z_{\odot}$).
\end{itemize}

\subsection{Filters}\label{sec:i.filters}
\begin{figure}
\centering
\includegraphics[width=20pc]{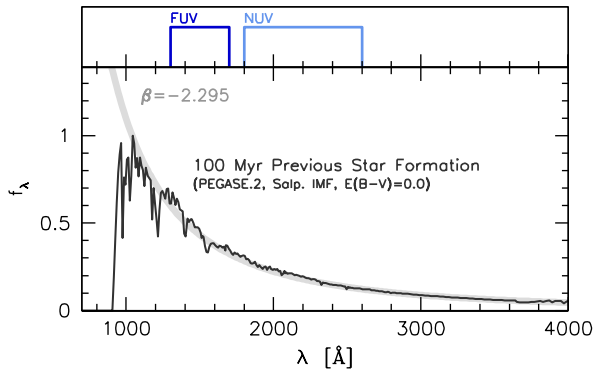}
\caption{The synthetic UV spectrum of a star forming galaxy ($100\,{\rm Myr}$ of previous continuous star formation, Salpeter IMF, $Z=Z_{\odot}$) produced with the {\sc pegase.2} population synthesis model (black line). The heavy grey line is a pure power law $f_{\lambda}\propto\lambda^{\beta}$ assuming the value of $\beta$ inferred from the FUV-NUV colour (transmission functions shown in upper panel). }
\label{fig:sp}
\end{figure}

In order to reproduce the techniques commonly applied to observations, we measure the UVC slope using two artificial broadband rest-frame filters (shown in Fig. \ref{fig:sp}), ${\rm FUV}$ ($T_{\lambda}=[0.13<\lambda/{\rm \mu m}<0.17]$\footnote{We utilise the Iverson bracket notation such that $[A]=1$ when $A$ is true and $0$ otherwise.}.) and ${\rm NUV}$ ($T_{\lambda}=[0.18<\lambda/{\rm \mu m}<0.26]$). These filters cover a similar wavelength range to the {\em GALEX} filter set. The use of rest-frame filters allows us to consistently compare the distribution of UVC slopes in galaxies at different redshifts.

\subsubsection{Parameterisation of the UVC slope}

Observations of the UVC colours of galaxies at different redshifts are difficult to compare as in each case the available filters probe different ranges of the UVC. To compare the UVC properties at different redshifts it is instead common to parameterise the continuum by a power law, with slope $\beta$ i.e.:
\begin{equation}
f_{\lambda}\propto\lambda^{\beta},
\end{equation}
or in terms of $f_{\nu}$,
\begin{equation}\label{eq:fnu}
f_{\nu}\propto\lambda^{2+\beta}\,.
\end{equation}
In the context of our choice of rest-frame filters, the conversion from $({\rm FUV-NUV})_{\rm AB}$ colour to $\beta$ is:
\begin{equation}\label{eq:beta_col}
\beta=C\times ({\rm FUV-NUV})_{\rm AB}-2\,,
\end{equation}
where $C$ is a conversion factor determined by convolving the two filter transmission functions with a power law spectrum. The value of $C$ is sensitive to the individual filters\footnote{For a combination of observer frame filters $C$ is also sensitive to the source redshift.} and for the $FUV$ and $NUV$ filters defined above is $C=2.49$. Fig. \ref{fig:sp} shows both the synthetic UV spectrum of a star forming galaxy and the power law using the value of $\beta$ inferred from the ${\rm FUV-NUV}$ colour. 

Throughout this work we make a distinction between the {\em intrinsic} UVC slope $\beta_{i}$ and the observed slope $\beta_{o}$. The intrinsic slope is assumed to be the slope produced by stars in the absence of any extrinsic effects such as dust extinction or nebular emission.

\section{Physical Properties Affecting the Observed UV Continuum Slope}\label{sec:pp}

The observed ultraviolet continuum slope of a galaxy is affected by various properties both intrinsic and extrinsic to the stars producing the UV emission. 

The intrinsic UVC of actively star forming galaxies is dominated by emission from the most massive, high-temperature stars ($m>2\,{\rm M_{\odot}}$, $OBA$ class). The UV ($1216\to 3000{\rm \AA}$) spectral flux density of these stars is well characterised by a power law with a {\em blue} slope (i.e. if $f_{\nu}\propto \lambda^{2+\beta}$ then $\beta<-2$) as shown in Fig. \ref{fig:sp}. This power-law behaviour principally arises because the UV continuum probes the Rayleigh-Jeans tail of the black body spectral energy distribution (though it deviates from a pure Rayleigh-Jeans behaviour because of the effects of opacity in the star's atmosphere) as shown by the synthetic spectrum in Fig. \ref{fig:sp}. At lower temperatures, the range $1216\to 3000{\rm \AA}$ probes  where the black body distribution peaks. This effect, combined with variation in the opacity (as a function of temperature etc.), results in the SEDs of the cooler, lower-mass stars having a redder UV continuum than those of stars with higher temperatures/masses. The sensitivity of the UVC of a star to its temperature (and opacity), and thus metallicity, mass and age means that the intrinsic continuum of a composite stellar population is sensitive to the distribution of stellar masses, ages and metallicities. The distribution of masses is in turn determined by the star formation history and IMF; thus both these factors affect the {\em intrinsic} UV continuum colours. 

In addition to these intrinsic properties the {\em observed} UVC is also affected by the presence of intervening dust through scattering and absorption and (typically to a much lesser extent) continuum emission from hot ionised gas. 

In the following sections (\S 3.1-3.4) we both outline the effect of dust (\S 3.1) and investigate the effect of intrinsic properties (including the star formation history, metallicity and IMF) on the observed UVC colours/slope.

\subsection{Dust}

The principal property which causes the observed UVC slope $\beta_{o}$ of luminous star forming galaxies to differ from the intrinsic slope is dust. Both empirically determined (e.g. Cardelli et al. 1989, Calzetti et al. 2000 - hereafter C00\footnote{The Calzetti et al. (2000) {\em starburst} attenuation curve encapsulates the effects of scattering and absorption while the Cardelli et al. (1989) curve describes extinction by a foreground screen and includes the prominent bump at $\lambda =2175{\rm\AA}$.}), and theoretically motivated dust attenuation laws suggest that attenuation increases rapidly from the NIR/optical through the UV. As such, the UV colour is potentially a powerful diagnostic of dust attenuation as the presence of dust will redden the observed slope compared to its intrinsic value. The exact relation depends on the characteristics of the attenuation curve in the UV; for example, depending on the choice of filters and redshift, a curve including the bump feature at $\lambda =2175{\rm\AA}$ (such as the Cardelli et al. 1989 extinction curve) implies a much weaker relationship between the UVC slope and dust than the C00 attenuation curve which decreases monotonically with wavelength. 

Here we use the C00 attenuation curve to see the effect of dust on the observed UVC slope. Assuming the underlying intrinsic UVC is described by a power law (Eq. \ref{eq:fnu}) a relationship between the observed and intrinsic UVC slopes, $\beta_{o}$ and $\beta_{i}$ respectively, and the attenuation in the FUV, $A_{\rm fuv}$, can be derived (c.f. C00),
\begin{equation}\label{eq:abeta}
A_{\rm fuv}=2.37\times[\beta_{o}-\beta_{i}].
\end{equation}
Assuming a synthetic intrinsic UVC (produced in this example using the {\sc pegase.2} SPS model) corresponding to $100\,{\rm Myr}$ previous constant star formation, and a Salpeter IMF with solar metallicity (as shown in Fig. \ref{fig:sp}) yields an intrinsic UVC slope of $\beta_{i}\simeq -2.3$. Inserting this into Eq. \ref{eq:abeta} yields
\begin{equation}\label{eq:auvbeta}
A_{\rm fuv}=2.37\beta_{o}+5.5,
\end{equation}
which is similar to the relation proposed by Meurer et al. (1999). This can be rewritten to express the uncertainty on the attenuation arising from the uncertainty in the intrinsic UVC slope,
\begin{equation}\label{eq:abetaerr}
\delta A_{\rm fuv}=2.37\times\delta\beta_{i}.
\end{equation}
This highlights the strong dependence of the inferred UV attenuation on the intrinsic UVC slope.

\subsection{Star Formation History}\label{sec:pp.sfh}

The strong variation with mass of the main-sequence lifetimes of stars means that the contemporary mass distribution, and thus UV continuum slope, is sensitive to a galaxy's star formation history. After less than $10\,{\rm Myr}$ of continuous star formation, a stellar population would still retain most of its original high-mass stars (i.e. up to this time, the contemporary mass distribution is similar to the IMF). However, after prolonged periods of star formation some fraction of the most massive stars will have evolved off the main-sequence. This reduces the relative contribution of these stars to the UV continuum, resulting in a redder slope (i.e $\beta>-2$). 

\begin{figure}
\centering
\includegraphics[width=20pc]{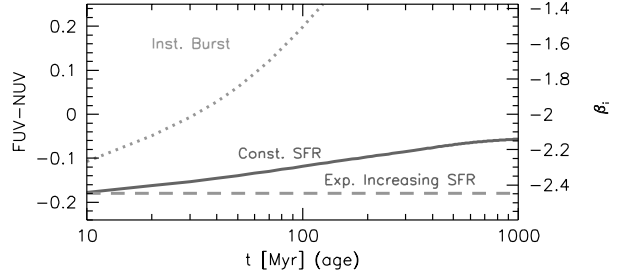}
\caption{The effect of the duration and form of previous star formation on the UVC colour. The three lines show the evolution of the UVC colour (left) and slope (right) of a stellar population forming stars at a constant rate (solid line), an exponentially increasing rate (dashed line) and as an instantaneous burst (dotted line).}
\label{fig:age}
\end{figure}

The limiting cases of the effect of the previous star formation history are shown in Fig. \ref{fig:age}. Here the evolution of the UVC slope is shown for three cases, an instantaneous burst, a constant SFH and an exponentially increasing SFH (assuming solar metallicity and a Salpeter IMF). The last case essentially leaves the slope constant, as the UVC is continuously dominated by the most massive $OB$-type stars. In the first case, the intrinsic UVC slope quickly reddens as it becomes dominated by progressively cooler stars. For the intermediate case, corresponding to a constant star formation rate, the intrinsic UVC slope evolves mildly, changing by only $\delta\beta_{i}\simeq 0.3$ between star formation durations of $10\,{\rm Myr}$ and $1\,{\rm Gyr}$. 

\begin{figure}
\centering
\includegraphics[width=20pc]{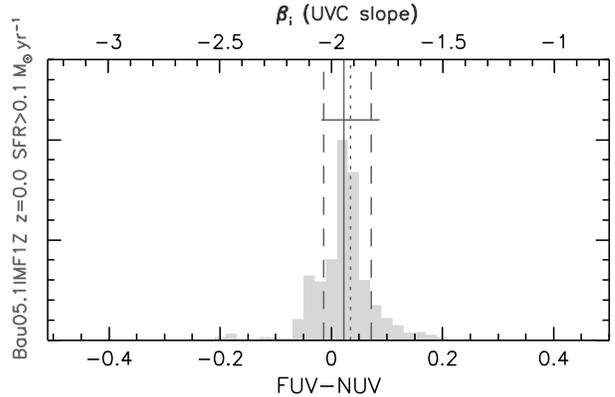}
\caption{The simulated distribution of intrinsic UVC slopes (top axis) and colours (bottom axis) for star forming galaxies (defined as those with SFR$>0.1\,{\rm M_{\odot}yr^{-1}}$) in the {\sc galform} model at $z=0$, assuming a single {\em constant} metallicity and universal IMF. The solid, dotted and dashed vertical lines denote the weighted mean, median and the $15.87$th to $84.13$th percentile range respectively. The solid horizontal line shows the standard deviation around the median. This, and subsequent, histograms are normalised by the modal value.}
\label{fig:d.SFR}
\end{figure}

Fig. \ref{fig:age} highlights that the UVC slope is sensitive to the recent star formation history. Estimating the uncertainty on the intrinsic UVC slope, $\delta\beta_{i}$, requires knowledge of the range of star formation histories. To obtain these we use simulated galaxies taken from the {\sc galform} model. In order to investigate the effect of the star formation history on the UVC we consider special cases in which the model is run with a fixed (solar) metallicity and a single IMF (Kennicutt 1983), with the variation in the intrinsic UVC slope then being entirely due the effect of the star formation history. The resulting distribution of UVC slopes, for galaxies selected as star forming (i.e. with star formation rates $\psi>0.1\,{\rm M_{\odot}yr^{-1}}$), is shown in Fig. \ref{fig:d.SFR}. The structure of this distribution is asymmetric with a long red tail. The asymmetry in the distribution (and the long red tail) is driven by the fact that while it is possible to obtain extremely red UVC slopes by having a dominant old stellar population, it is impossible to push the UVC slope to be arbitrarily blue. Despite the non-gaussian structure of this distribution, the standard deviation around the median provides a close indication of the $15.87$th to $84.13$th percentile confidence interval. The standard deviation of this distribution is $\sigma_{\beta}\simeq 0.13$; assuming the formula relating the uncertainty on $\beta_{i}$ to the UV attenuation (Eq. \ref{eq:abetaerr}), this suggests an intrinsic uncertainty on the UV attenuation due to the variation in the SFH of the galaxy population of $\delta A_{\rm fuv}\simeq 0.31$.

\subsection{Metallicity}\label{sec:pp.Z}

\begin{figure}
\centering
\includegraphics[width=20pc]{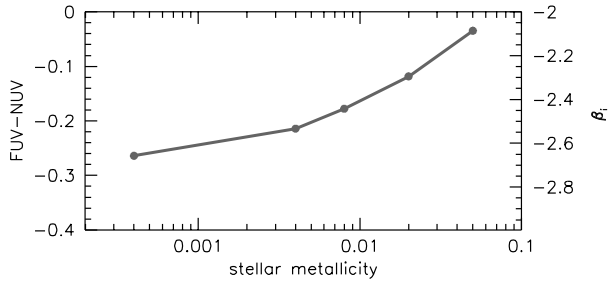}
\caption{The effect of assuming different values of the stellar metallicity on the intrinsic UVC colour (left) and slope (right) derived from the {\sc Pegase.2} SPS model (assuming assuming a $100\,{\rm Myr}$ previous duration of continuous star formation prior to measurement and a Salpeter IMF).}
\label{fig:C_Z}
\end{figure}

Both the opacities and effective temperatures, and thus the ultraviolet continua, of stars are affected by their chemical composition; stars with lower metal abundances typically have bluer UVC slopes. Hence, the intrinsic UVC slope is also sensitive to the metallicity of the UV luminous population, or specifically to the {\em recent} metal enrichment history. Fig. \ref{fig:C_Z} shows the impact of metallicity on the intrinsic UVC slope (assuming constant star formation in the $100\,{\rm Myr}$ prior to observation and a Salpeter (1955) IMF, using the {\sc pegase.2} population synthesis model); changing the metallicity from $Z=0.02$ to $0.004$ causes the intrinsic UVC slope $\beta_{i}$ to steepen by $\simeq 0.25$. Thus the UV attenuation, $A_{fuv}$, of a galaxy which has $Z=0.004$, but which is assumed to have $Z=0.02$ would be overestimated by $\simeq 0.6\,{\rm mag}$ due to the difference in the intrinsic UVC colour.   

\begin{figure}
\centering
\includegraphics[width=20pc]{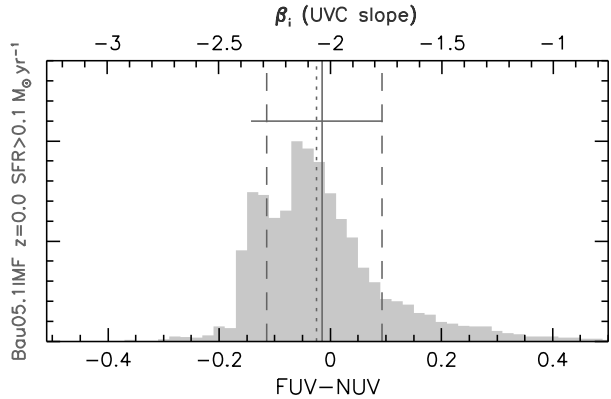}
\caption{The simulated distribution of UVC colours for star forming galaxies (SFR$>0.1\,{\rm M_{\odot}yr^{-1}}$) at $z=0$ assuming a single initial mass function but varying metal enrichment and star formation history. The solid, dotted and dashed vertical lines denote the weighted mean, median and $15.87$th and $84.13$th percentile range respectively. The solid horizontal line shows the standard deviation around the median.}
\label{fig:d.SFRZ}
\end{figure}

As with the SFH, accounting for the metallicity effects is difficult without some knowledge of the metal enrichment history. Once again we turn to the {\sc galform} model to provide plausible distributions of star formation and metal enrichment histories from which we can determine the intrinsic UVC slope. Including the predicted variation in metal enrichment histories of galaxies widens the intrinsic distribution of UVC slopes compared to that obtained assuming a single metallicity (the resulting distribution is shown in Fig. \ref{fig:d.SFRZ}). The standard deviation of the intrinsic UVC slope distribution increases to $\sigma_{\beta}\simeq 0.30$ compared with $\sigma_{\beta}\simeq 0.13$ with no metallicity variation. Using Eq. \ref{eq:abetaerr}, this suggests an intrinsic uncertainty in the UV attenuation inferred from the UVC colour of $\delta A_{\rm fuv}\simeq 0.7$. This suggests that, even in the absence of photometric noise or redshift uncertainties, the dust attenuation of a single object can not be measured more accurately than $\delta A_{\rm fuv}\simeq 0.7$ without additional information to constrain the star formation history or metallicity. 

\subsection{Initial Mass Function}\label{sec:pp.imf}

The IMF ($\xi(m)={\rm d}N/{\rm d}m$) describes the stellar mass distribution of a single stellar population with $t_{\rm age}=0$. A range of parameterisations of the IMF exist, with the simplest taking the form of single power law (i.e. $\xi(m)\propto m^{\alpha}$), e.g. the Salpeter (1955) IMF with $\alpha=-2.35$. Updated parameterisations take into account the observed flattening of the IMF below some characteristic mass (e.g. $m_{c}\simeq 1\,{\rm M_{\odot}}$ for the Kennicutt IMF) by adopting a broken power law parameterised by $\alpha_{1}$ and $\alpha_{2}$ (the low and high-mass slopes respectively). 

\begin{figure}
\centering
\includegraphics[width=20pc]{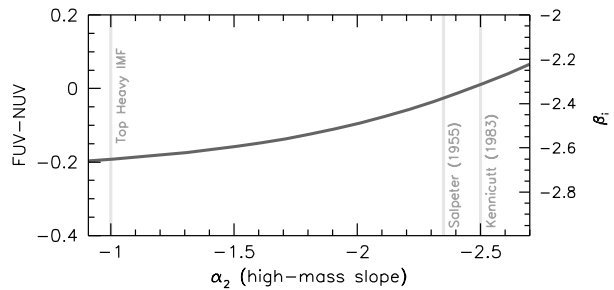}
\caption{The effect of the choice of the high-mass slope $\alpha_{2}$ of the IMF on the UVC colour/slope derived from the {\sc Pegase.2} SPS mode (assuming assuming a $100\,{\rm Myr}$ previous duration of continuous star formation and solar metallicity).}
\label{fig:C_a2}
\end{figure}

The IMF can be modified in a number of ways in the context of this parameterisation. One method is to simply vary the high-mass slope $\alpha_{2}$. Steepening the high-mass slope (decreasing $\alpha_{2}$) will reduce the relative contribution of very-high mass stars to the UV luminosity; this will have the effect of reddening the UVC colour/slope. This can be seen in Fig. \ref{fig:C_a2} (assuming $100\,{\rm Myr}$ previous duration of continuous star formation and solar metallicity); flattening $\alpha_{2}$ from $-2.35\to -1.5$ reduces the slope by $\Delta\beta_{i}\simeq 0.2$ assuming the same star formation and metal enrichment history. 

The effect of the IMF is potentially important as a number of studies have suggested the high-mass slope of the IMF may be flatter than Salpeter (e.g. Wilkins et al. 2008b) or that the IMF may {\em effectively} vary from galaxy to galaxy (e.g. Baldry \& Glazebrook 2003, Wilkins et al. 2008a, Hoversten \& Glazebrook 2008, Lee et al. 2009, Finkelstein et al. 2011). 

\begin{figure}
\centering
\includegraphics[width=20pc]{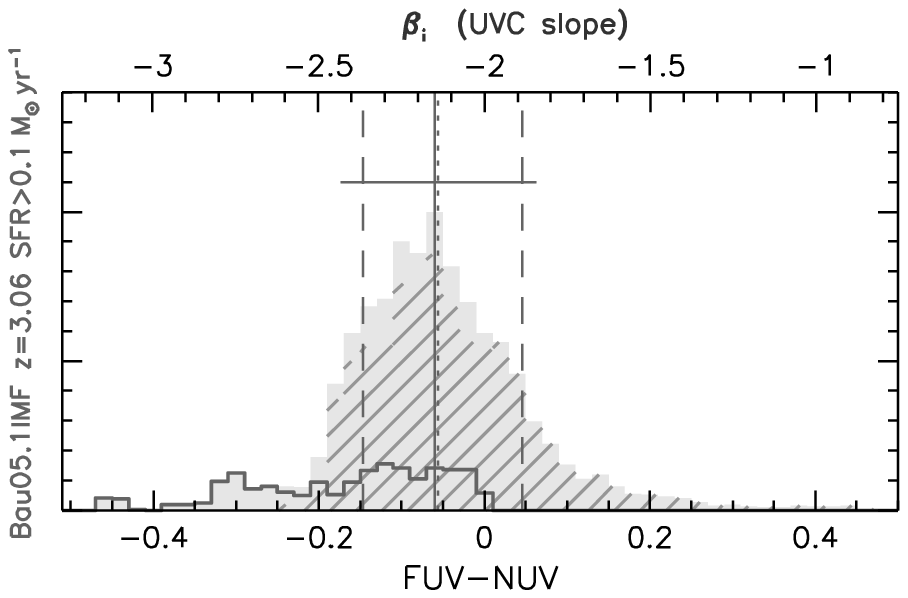}
\includegraphics[width=20pc]{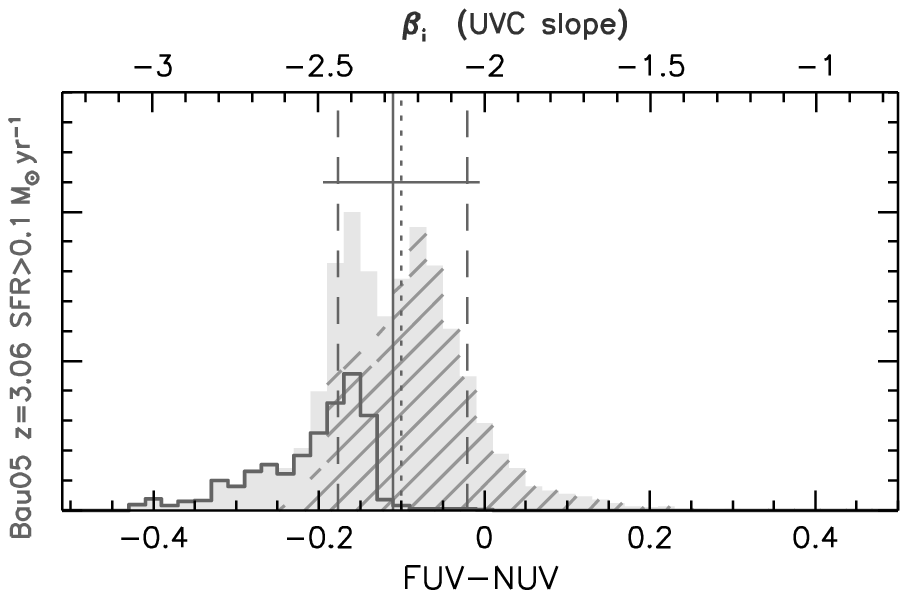}
\caption{The simulated distribution of intrinsic UVC colours and slopes for star forming galaxies ($\psi>0.1\,{\rm M_{\odot}yr^{-1}}$) at $z=3$ assuming the B05 model with a single Kennicutt (1983) IMF (top) and the default implementation with two IMFs (bottom). The histogram enclosed by the dark line shows the distribution for galaxies whose total SF is dominated by the burst mode while the hatched histogram shows those dominated by quiescent star formation (the grey shaded histogram is for all star forming galaxies).}
\label{fig:a2dist}
\end{figure}

The {\em default} implementation of the B05 {\sc galform} model adopts a top-heavy IMF ($\xi\propto m^{-1}$) in merger triggered SF, while retaining the Kennicutt IMF in quiescent SF. At low redshift, where our attention has been focused thus far, the fraction of star formation occurring with the flat, top-heavy IMF is small in the B05 model. This fraction increases with redshift such that at $z\simeq 3$ the contribution to the star formation rate density from each mode is roughly similar. In Fig. \ref{fig:a2dist} we compare the intrinsic UVC slope distribution of galaxies at $z\simeq 3$ using both the default implementation of B05 model (i.e. with the top-heavy IMF mode of star formation in bursts) and a variant in which a universal IMF is used. The main difference between the two is that the distribution of UVC colours in the default B05 model is on average bluer and the width of the distribution slightly smaller. A second peak also emerges associated with burst-driven SF. This can be seen clearly in Fig. \ref{fig:a2dist} where the distribution for galaxies whose total star formation is dominated by the burst mode is shown alongside the full distribution.

\section{Trends with SFR, UV Luminosity, Stellar Mass and Redshift}\label{sec:a}

In the previous section we showed that the distribution of intrinsic UVC slopes (of star forming galaxies) is affected by the IMF, and the star formation and metal enrichment histories of galaxies. In this section we investigate if the median intrinsic UVC slope correlates with a galaxy's star formation rate, intrinsic UV luminosity, stellar mass and redshift. A correlation of the intrinsic slope with any of these properties may result in an erroneous observed correlation between the property and the inferred dust extinction, or may reinforce a weak or negligible correlation. 

\begin{figure*}
\centering
\includegraphics[width=40pc]{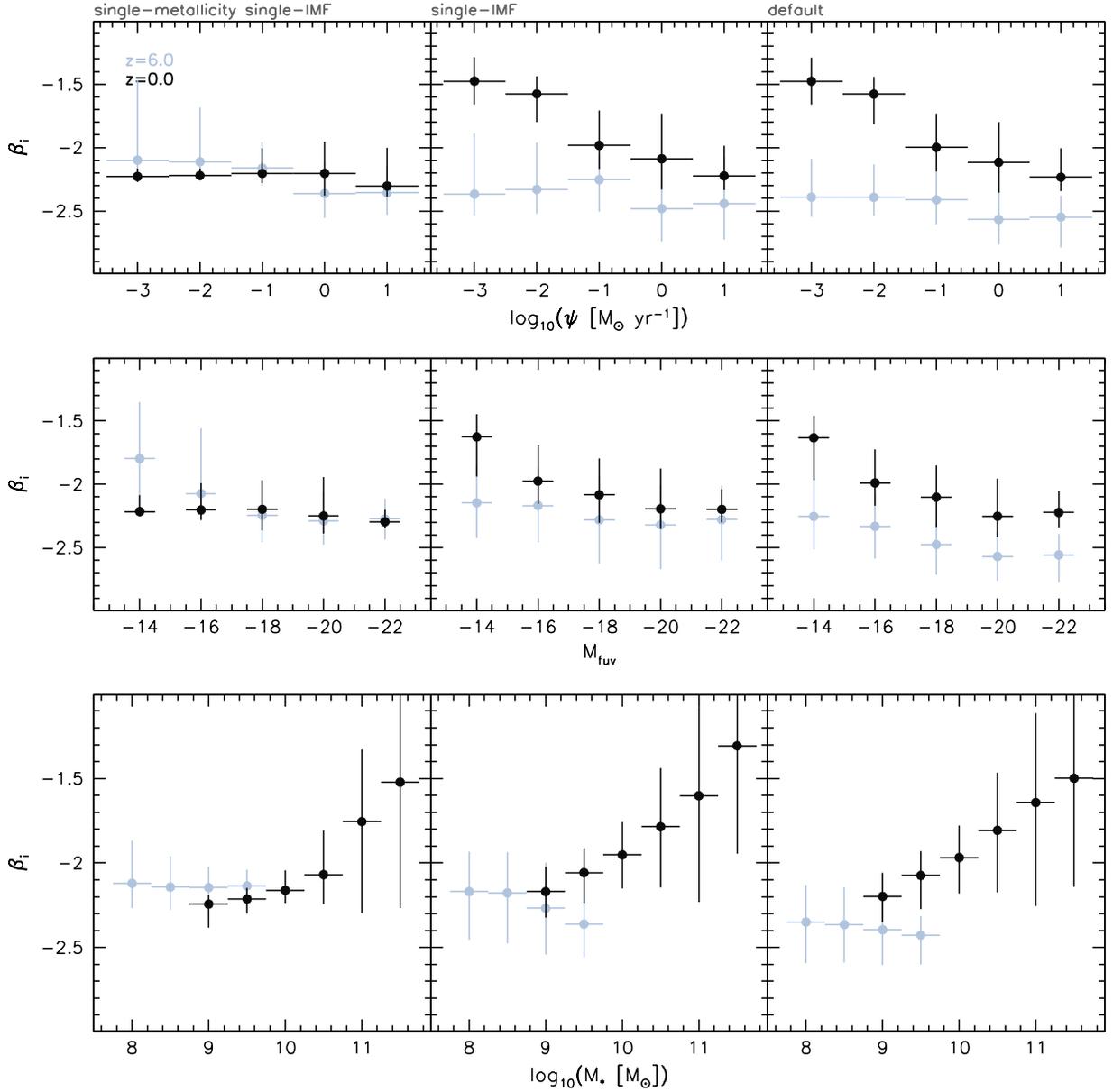}
\caption{The median intrinsic UVC slope binned by star formation rate (top), intrinsic FUV absolute magnitude (middle) and stellar mass (bottom) at $z=0$ (black) and $z=6$ (grey) for each of the three model variants: single-metallicity and single-IMF (left), single-IMF (middle) and {\em default} (right). In each case the horizontal bar denotes bin width while the vertical bar denotes the $15.87{\rm th}$ to $84.13{\rm th}$ percentile range of the {\sc galform} model predictions.}
\label{fig:trend1}
\end{figure*}

Fig. \ref{fig:trend1} shows the median intrinsic UVC slope for the three variants of the B05 model (single IMF and metallicity, single IMF, and {\em default}) binned by the star formation rate, intrinsic UV luminosity and stellar mass, which we discuss in turn below.

\subsection{Star Formation Rate}\label{sec:a.sfr}

The top panels of Fig. \ref{fig:trend1} show the median UVC slope (and $15.87{\rm th}$ to $84.13{\rm th}$ percentile range) for galaxies binned by their star formation rate at $z=0$ and $z=6$ using the three variants of the B05 model (single-metallicity and single-IMF, single-IMF and {\em default}). For the single-metallicity single IMF model there is no correlation between the star formation rate and median intrinsic UVC slope, irrespective of redshift. However for both the single-IMF and {\em default} variants of the model there is a negative correlation at low-redshift. This trend becomes progressively flattened at higher-redshift and for $z>4$ is essentially flat ($z=6$ is shown in Fig. \ref{fig:trend1}). The existence of this trend in the single-IMF and {\em default} variants of the model and its absence in the single-metallicity and single-IMF model suggests this trend is driven by the metallicity variation as a function of star formation rate. At high-redshift the variation in metallicity is less pronounced leading to the absence of any correlation.

\subsection{Intrinsic UV Luminosity}\label{sec:a.L_fuv}

Related to the star formation rate but more useful in an observational context is the intrinsic FUV luminosity, $L_{\rm fuv}$. The middle panels of Fig. \ref{fig:trend1} shows the median intrinsic UVC slope binned by the intrinsic absolute FUV magnitude. Due to the close connection between the SFR and the intrinsic UV luminosity a similar trend exists to that described above; at low-redshift there is a mild trend such that more luminous galaxies have an intrinsically bluer UVC colour (in the single-IMF and {\em default} variants of the model), while at higher-redshift the correlation is virtually flat. The flat trend at high-redshift, reinforces the conclusions of several studies (e.g. Wilkins et al. 2011b) that galaxies with brighter {\em observed} UV luminosities have greater dust attenuation than their low luminosity counterparts. 

\subsection{Stellar Mass}\label{sec:a.smass}

The lower panel in Fig. \ref{fig:trend1} shows the variation of $\beta$ with stellar masses for galaxies around $M^{*}$ at $z=0$ and $z=6$. At $z=0$ (but extending to $z\simeq 1$), and for all three model variants, the median UVC colour is also correlated with stellar mass, becoming redder at high-stellar masses. The width (as measured by the confidence interval) of the UVC slope distribution also increases dramatically to high stellar masses. The fact that this trend exists for all three models suggests that it is driven predominantly by the star formation history in so far as more massive galaxies are typically composed of older populations (e.g. Kauffmann et al. 2003). At higher-redshift this correlation flattens, to the extent that at $z=6$, the trend is virtually flat. 

One implication of the strong correlation between stellar mass and UVC slope is that assuming a constant intrinsic slope to convert the observed slope to an attenuation may introduce an artificial correlation. As an example, applying the same intrinsic slope found at $\log_{10}(M_{*}/M_{\odot})=9$ ($\beta_{i}\simeq -2.2$) would result in a systematic over-estimation of the attenuation at $\log_{10}(M_{*}/M_{\odot})=11$ ($\beta_{i}\simeq -1.6$) of $\delta A_{fuv}\simeq 1.42$ assuming the single-IMF variant of the B05 model.

\subsection{Redshift}\label{sec:a.z}

\begin{figure}
\centering
\includegraphics[width=20pc]{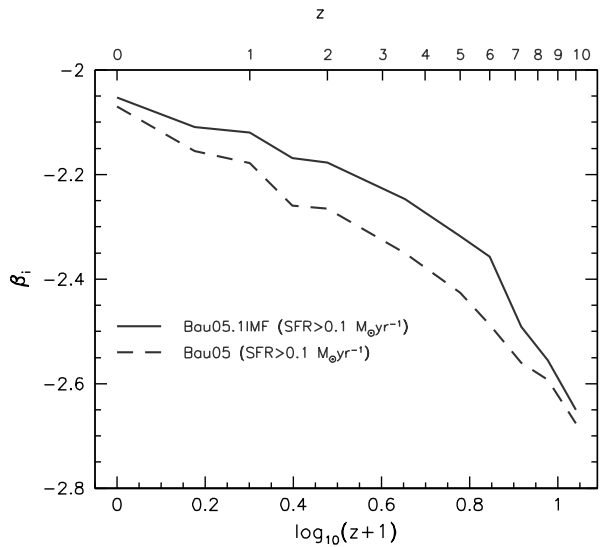}
\caption{The evolution of the median UVC slope of star forming galaxies as a function of redshift for both the single IMF and {\em default} implementation of the B05 model.}
\label{fig:z}
\end{figure}

Potentially the most important trend is the evolution of the median UVC slope with redshift. Shown in Fig. \ref{fig:z} is the redshift evolution of the median UVC slope for star forming galaxies (SFR$>0.1\, M_{\odot}yr^{-1}$) assuming both the single IMF (solid line) and {\em default} (dashed line) implementations of the B05 model.

In both cases the median intrinsic UVC slope evolves, becoming progressively redder towards lower redshift. The cause of this is a combination of both the increasing average metallicity of recently formed stars and the changing stellar mass distribution in star forming galaxies. As progressive generations of stars form, the mass distribution in galaxies becomes increasingly dominated by low and intermediate mass stars with long main-sequence lifetimes. The evolution of the median UVC slope assuming the {\em default} implementation of the B05 model is similar to the single-IMF implementation of the B05 model, though slightly bluer over the entire history. Specifically, at $z>1$ the difference between models is significant ($\delta\beta_{i}\simeq 0.1$) while at lower-redshift ($z<0.5$) the difference becomes negligible $\delta\beta_{i}<0.05$, reflecting the smaller contribution of merger-driven star formation to the total star formation rate density at these epochs. 

If this decrease in the median intrinsic UVC slope is not properly taken into account, and, for example, a relation relevant at low-redshift is applied to high-redshift galaxies (as is typically done in the literature), the UV attenuation inferred will be systematically underestimated. For example, applying a relation based on the intrinsic UVC slope distribution at $z=0$ to galaxies at $z=7$ would result in the systematic under-estimation of the UV attenuation by $\delta A_{fuv}\simeq 0.97$ and the possible introduction of an erroneous redshift trend or at least an over-estimation of the redshift evolution. 

In addition, in both cases, the width of the distribution also evolves, becoming slightly narrower at higher redshift (at $z=6$ $\sigma_{\beta}=0.26$, c.f. $z=6$ $\sigma_{\beta}=0.30$ for the single-IMF implementation). For the default implementation we also see the emergence of bimodality in the distribution as a result of the merger-driven SF occurring with a top-heavy IMF. At low redshift, where the contribution of merger-driven star formation is small, and at very-high redshift, where the contribution of quiescent star formation is low, the bimodality essentially disappears.

\section{Conclusions}\label{sec:c}

We have used realistic star formation and metal enrichment histories predicted by the {\sc galform} semi-analytical galaxy formation model to investigate the intrinsic distribution of ultraviolet continuum (UVC) colours and slopes. We find, for star forming galaxies at low-redshift, that the standard deviation of the distribution of UVC slopes (parameterised by the power-law index $\beta_{i}$) is $\sigma_{\beta}\simeq 0.30$ (assuming a single IMF). Assuming the Calzetti et al. (2000) reddening curve this suggests an intrinsic uncertainty of $\delta A_{1500}\simeq 0.71$ in the UV attenuation inferred from the UVC slope for an individual object without any aditional information to constrain the star formation or metal enrichment histories. 

We also investigated how the median intrinsic UVC slope and the width of the $\beta_{i}$ distribution correlates with various properties including the star formation rate, intrinsic UV luminosity, stellar mass and redshift. At low-redshift we find that the median UVC slope is sensitive to the star formation rate, intrinsic UV luminosity and stellar mass, with more massive galaxies typically having redder UV continuum slopes and those with higher star formation rates (and higher UV luminosities) typically being bluer. At higher-redshift these correlations flatten and become less important. The model also suggests significant evolution of median UVC slope with redshift (though the width of the distribution changes only mildly). These various correlations suggest that trends of the observed UVC slope with stellar mass or redshift are not driven entirely (if at all) by dust but partly by the evolution of the star formation and metal enrichment histories of galaxies. In order to correctly interpret observations, in the context of dust reddening, it is then important to compare the observed distribution of UV continuum slopes with a model including a physically motivated treatment of dust but taking also into account the predicted intrinsic distribution. In a follow up paper (Wilkins et al. 2012) we compare the observed distribution with the predicted intrinsic one for bright $z\approx 3-5$ star forming galaxies.

\subsection*{Acknowledgements}

We would like to thank the anonymous referee for helpful suggestions and comments which improved the quality of this paper. The calculations for this paper were performed on the ICC Cosmology Machine, which is part of the DiRAC Facility jointly funded by STFC, the Large Facilities Capital Fund of BIS, and Durham University. SMW acknowledges support from STFC. VGP acknowledges support from the UK Space Agency. VGP, CGL, \& CMB acknowledges support from the Durham STFC rolling grant in theoretical astronomy.

\bsp

\end{document}